\documentclass[aps,prl,twocolumn,showpacs,groupedaddress]{revtex4}  % for review and submission 
\usepackage{graphicx}  % needed for figures
\usepackage{dcolumn}   % needed for some tables
\usepackage{bm}        % for math
\usepackage{amssymb}   % for math
\usepackage{epstopdf}

\newcommand{\textss}[1]{\scriptsize \mbox{#1}}

\newcommand{\Epsilon}{\mathcal{E}}

\begin{document}

\title{Extreme Nonlinear Optics in a Femtosecond Enhancement Cavity}

\author{T. K. Allison, A. Cing\"{o}z, D. C. Yost, and J. Ye }
\affiliation{JILA, National Institute of Standards and Technology and Unviversity of Colorado, Boulder, CO 80309}
\begin{abstract}

Intrinsic to the process of high-order harmonic generation is the creation of plasma and the resulting spatiotemporal distortions of the driving laser pulse.   Inside a high-finesse cavity where the driver pulse and gas medium are reused, this can lead to optical bistability of the cavity-plasma system, accumulated self-phase modulation of the intracavity pulse, and coupling to higher order cavity modes. We present an experimental and theoretical study of these effects and discuss their implications for power scaling of intracavity high-order harmonic generation and extreme ultraviolet frequency combs. 
\end{abstract}

\maketitle

%A laboratory scale XUV and soft x-ray light source can be realized through high order harmonic generation (HHG) \cite{Brabec_RMP2000}. HHG transfers the spatial and temporal coherence and short pulses of femtosecond lasers to the soft x-ray and XUV and is now being used for many exciting applications. 

High-order harmonic generation (HHG) is initiated through field ionization of a gas medium,  and thus high peak laser intensities are needed. This is typically realized with low repetition rate ($< 100$ kHz) amplified femtosecond laser systems producing high energy pulses $( > 100$ $\mu$J), with average powers up to tens of Watts. In an alternative method, the modes of a frequency comb and a high-finesse external cavity can be locked, leading to $ > 100$ MHz repetition rates and multi-kW average powers, while still supporting peak intensities sufficient for HHG \cite{Jones_PRL2005,Gohle_Nature2005}. Efficient conversion of such high power to the extreme ultraviolet (XUV) holds great promise as a route to table top, high average brightness sources of coherent XUV light. Such a source could have a large scientific impact, particularly in applications where high repetition rate and/or temporal coherence is demanded, such as XUV frequency metrology \cite{Kandula_PRL2010} and time resolved photoemission \cite{Rohwer_Nature2011} or photoionization coincidence \cite{Sandhu_Science2008} spectroscopies. 

Since the first demonstrations in 2005 \cite{Jones_PRL2005,Gohle_Nature2005}, much progress has been made in the construction of high power frequency combs \cite{Ruehl_OptLett2010,Cingoz_OptLett2011} and special optics for the efficient output-coupling of the generated HHG light from the cavity \cite{Yost_OptLett2008,Yang_OptExp2011}. Recently, intracavity powers as high as 18 kW have been reported while maintaining femtosecond pulses \cite{Pupeza_OptLett2010}. However, while plasma refraction has long been known to be a critical issue in the phase matching of HHG, a cavity enhanced system operates with the additional constraints that the plasma must not ruin the cavity finesse or destroy the delicate resonance between comb and cavity. The effects of the plasma on the cavity performance have remained open questions and are the subject of this Letter. Only with these understandings have we been able to improve the cavity HHG yield by more than an order of magnitude and report here record levels of HHG performance in the XUV. Furthermore, understanding and controlling extreme-nonlinear-optical modulation of the intracavity light is critical for the phase coherence of the generated XUV frequency comb, because phase modulation of the fundamental appears multiplied by the harmonic order in the XUV light. Indeed, limited knowledge of the plasma phase shifts inherent to HHG was the dominant source of systematic error in a recent demonstration of XUV spectroscopy \cite{Kandula_PRL2010}.

The propagation of femtosecond pulses in a field-ionizing gas medium has been previously studied theoretically and experimentally \cite{Geissler_PRL1999, Penetrante_JOSAB1992}. The laser field loses energy to the production of photoelectrons and experiences a rapid ramp in the index of refraction on the femtosecond time scale of the pulse, resulting in self-phase modulation and a frequency shift to the blue. Also, due to the high nonlinearity of field ionization, the plasma generated by a focused laser beam is far from spatially uniform, and acts as a negative lens, defocusing the beam. In a high-finesse cavity, the strength of these interactions between pulse and medium is effectively increased for two reasons. First, because the intracavity pulse is reused, pulse and wavefront distortion accumulates over many round trips. Second, if the repetition rate is sufficiently high, neither plasma expansion \cite{Kanter_JApplPhys2008} nor gas jet flow is fast enough to clear the plasma from the focal volume and replenish it with neutral atoms within one cavity round trip. The intracavity pulse then gets an additional phase shift due to the residual plasma from previous round trips. In these ways, the plasma acts as a highly nonlinear optical element in the cavity, with responses on femtosecond and nanosecond time scales, presenting a novel and nontrivial problem in nonlinear optics.

Our general cavity layout has been described previously in Ref. \cite{Yost_OptLett2008}. The system is pumped with a Yb$^+$ fiber comb system capable of delivering up to 80 W of average power in 120 fs pulses with a repetition rate $f_{\textss{rep}} = 154$ MHz and a center wavelength $\lambda_c$ = 1070 nm \cite{Ruehl_OptLett2010}. Xe is introduced at the intracavity focus with a 200 $\mu$m diameter glass nozzle. The focus size (FWHM) was estimated to be 29 $\mu$m (horizontal) $\times$ 17 $\mu$m (vertical) from an ABCD matrix analysis of the cavity and inspection of the frequency spacings of the cavity's higher order modes. The cavity is strongly overcoupled \cite{Siegman_book1986}, with the round trip loss dominated by the 1.5$\%$ transmission of the input coupler, for a finesse of $\mathcal{F} \sim 400$ and an optimum power enhancement factor of $ \sim 250$. In practice, a lower power enhancement of 200-220 is observed due to imperfect mode matching. The intracavity power is measured with a photodiode monitoring the transmitted light  through a cavity mirror of calibrated transmission. Frequency-lock between the comb and cavity is accomplished using the Pound-Drever-Hall (PDH) technique \cite{Jones_OptLett2002}. The generated high order harmonics are coupled out of the cavity with a small period diffraction grating etched into the surface of a high reflector \cite{Yost_OptLett2008}. The 13$^{\textss{th}}$ harmonic power is measured with a Si XUV photodiode coated with indium (IRD Inc. AXUV100In/MgF$_2$), and was confirmed to be insensitive to scattered IR and 3$^{\textss{rd}}$ harmonic light in the present geometry. 

\begin{figure}[t]
\begin{center}
\includegraphics[width = 9.0 cm]{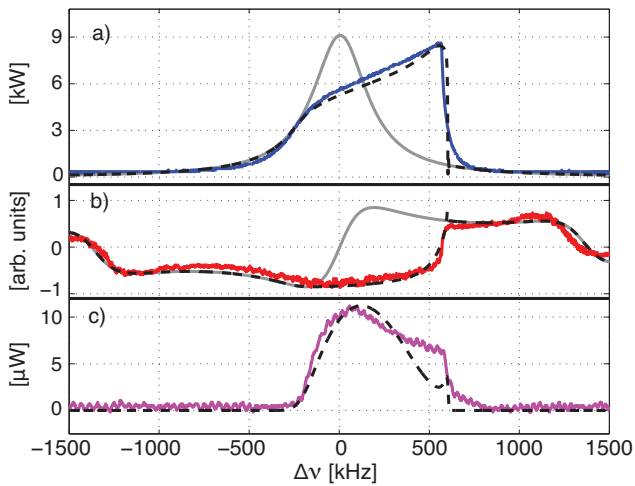} 
\caption{a) 
Intracavity power, b) PDH error signal, and c) 13th harmonic power for a comb-cavity detuning scan with 40-W pump power. The solid colored lines are experimental data and the black dashed lines are model results. The solid gray lines illustrate the intracavity power and error signal in the absence of gas.}
\label{sweeps_fig}
\end{center}
\end{figure}

Resonance between the pump comb light and the optical cavity is achieved when the optical frequencies of the comb lines are matched to the respective optical resonances of the cavity. The nonlinear response of the system is clearly observed by sweeping the optical detuning $\Delta \nu \equiv \nu_{\textss{comb}} - \nu_{\textss{cavity}}$ across this resonance condition, as shown in Fig. \ref{sweeps_fig}. To acquire the data, the carrier-envelope offset frequency of the comb is first matched to that of cavity, and then the cavity length is scanned across the resonance while the intracavity power, PDH error signal, and 13th harmonic power are recorded with a digital oscilloscope. The detuning is scanned from negative to positive $\Delta \nu$ in Fig. \ref{sweeps_fig} and the frequency axis is calibrated using the resonances of the PDH sidebands observed in the error signal at $\pm$ 1.3 MHz. In the absence of gas (gray curves), the intracavity power is an approximately Lorentzian function of detuning, and the PDH error signal, proportional to the first derivative of the line shape for small detuning, crosses zero at resonance. With gas flow and the onset of plasma creation, the intracavity power is reduced and the resonance condition is shifted to a blue detuning of more than a full cavity linewidth.

All of the aforementioned consequences of light propagation in a field-ionizing medium can contribute to the nonlinear response of the cavity. To unravel their relative contributions, we have developed a simple model appropriate for multicycle pulses under the present circumstances, in which the cavity finesse is high and the gas medium density-length product sufficiently low that the change in the intracavity pulse induced in one round trip is small. We use a frequency domain approach to describe the evolution of the intracavity comb on the time scale of tens of cavity round trips, denoted with variable $T$, and a time domain description to calculate the plasma ionization, loss, and high order harmonic generation on the femtosecond time scale, denoted with variable $t$. The expansion dynamics of laser produced Kr plasmas have been studied by Kanter et. al. \cite{Kanter_JApplPhys2008}, who found the centerline density to decay with a time constant of $\sim 20$ ns governed by the ion speed of sound. The time scale for jet flow to replenish the focal volume is similar, on the order of 70 ns. The plasma decay is then fast compared to dynamics of the intracavity comb, with a characteristic time scale of $\mathcal{F}/(\pi f_{\textss{rep}}) \sim 800$ ns, but slow on the femtosecond time scale of the pulse. Expressing the plasma density as $\rho_0\eta$, where $\rho_0$ is the atom number density and $\eta$ is the fraction of atoms that are ionized, the ionization fraction is then logically divided into two components: a steady-state component $\bar{\eta}(T)$ that effectively tracks the intracavity pulse intensity, and a dynamic component $\eta(t)$ responsible for self-phase modulation.

The intracavity electric field is described by a discrete set of $T$-dependent spectral elements, $A_j(T) \equiv A(T,\omega_j)$, each representing a portion of the intracavity spectrum with optical frequency $\omega_j$. The time evolution of the spectral elements is governed by a set of coupled differential equations:
\begin{equation}\label{AODE}
	\frac{dA_j}{dT} = \sqrt{\delta}E_j - \frac{1}{2}(\delta+\gamma)A_j + i(\theta_j + \phi_j)A_j + f_j(\left\{ A_k \right\}),
\end{equation}
where $\delta$ is the input coupler transmission, $\gamma$ is the round trip loss of the rest of the cavity, and $E_j = E(\omega_j)$ is a set of spectral elements describing the electric field of the pump light. $T$ is expressed in units of the cavity round trip time, $1/f_{\textss{rep}}$ = 6.5 ns. The $\theta_j$ describe the round trip phase shifts of the intracavity $A_j$ with respect to the pump $E_j$ due to the cavity detuning, analogous to $\Delta\nu$ of Fig \ref{sweeps_fig}. The differential equations (1) are coupled through the steady-state plasma phase shifts $\phi_j(A_1,...,A_N)$ and the function $f_j(A_1,...,A_N)$, which accounts for the round-trip self-phase modulation and loss. In this way, different portions of the intracavity comb interact nonlinearly via the plasma. Both $\phi_j$ and $f_j$ are calculated from the $t$-domain description of the intracavity pulse as described below.

%The computational speed and numerical stability of the $t$-domain calculations for the elements $f_i$, which must be repeated many times, are greatly improved if an approach using an electric field envelope function is used, as opposed to constructing a full oscillatory electric field.

While the sub-cycle dynamics of field ionization, with very high frequency content, are essential for understanding HHG, they are not critical for explaining the distortion of the fundamental pulse if the pulse is multicycle and the ionization per cycle is small. Accordingly, at each time step, ionization, self phase modulation, and loss are calculated using an electric field envelope $\Epsilon(t)$, calculated from the $A_j$ via discrete time Fourier transform (DFT). The empirical formula of Tong and Lin \cite{Tong_JPhysB2005} is used to construct cycle-averaged and peak ionization rates, $\bar{w}(\Epsilon)$ and $w(\Epsilon)$. The ionization fraction during the pulse, $\eta(t)$ is then calculated via
\begin{equation}
	\eta(t) =  1- \left[1-\bar{\eta}(T)\right]\exp \left[-\int_{-\infty}^{t} dt \mbox{ } \bar{w}(\Epsilon)\right]
\end{equation}
The steady-state ionization $\bar{\eta}(T)$ is calculated from a balance between the plasma creation per round trip, $\dot{\bar{\eta}} = \eta(t = \infty) - \bar{\eta}(T)$, and the simple decay law $\dot{\bar{\eta}} = -k_p\bar{\eta}$, where $k_p$ is a constant.

To calculate the changes in the field envelope upon one pass through the plasma, we introduce the following approximate expression derived from the propagation equation of Geissler \cite{Geissler_PRL1999}:
\begin{eqnarray}\label{Epsapprox}
	\Delta\Epsilon(t) & = & \alpha_1\Epsilon(t) \bigg\{ -i r_0\lambda_c\rho_0L\left[\eta(t)-\bar{\eta}(T)\right] \nonumber \\ 
             	&   &  -I_p\frac{2\pi \rho_0 L}{c}\frac{\left[1-\eta(t)\right]w(\Epsilon)}{|\Epsilon|^2}\bigg\}
\end{eqnarray}
where $L$ is the medium length, $I_p$ is the ionization potential of the target gas, $r_0$ is the classical radius of the electron, and $c$ is the speed of light. The first term on the right-hand side accounts for the dynamic change in the plasma refraction during the pulse. Note that the steady state component, already included in $\phi_j$ in equation Eq. 	(\ref{AODE}), has been subtracted off. The second term, essentially a low pass filtered version of the ionization loss term of Geissler \cite{Geissler_PRL1999}, accounts for the energy lost by the laser in ionizing the medium. Equation (\ref{Epsapprox}) neglects energy transferred to the electrons via their nonzero drift velocity, which is small compared to the ionization loss for Keldysh parameters of order unity or larger \cite{Geissler_PRL1999}. The coupling terms $f_j$ on the RHS of (\ref{AODE}) are calculated from the inverse DFT of $\Delta \Epsilon$. The reduction factor $\alpha_1 <1$ crudely accounts for the 3D effect that only the center of the Gaussian focus experiences self-phase modulation, so that the effective modulation averaged over the entire wavefront is less. We include a similar factor $\alpha_2 < 1$ in the calculation of $\phi_j$ from $\bar{\eta}(T)$ via
\begin{equation}
  \phi_j = -\alpha_2\bar{\eta} r_0 \lambda_c L \rho_0 \left[1-\Delta\omega_j/\omega_c + (\Delta\omega_j/\omega_c)^2 - ... \right]
\end{equation}
where $\Delta \omega_j$ is the optical angular frequency difference from the carrier frequency $\omega_c$. For comparison with experiment, an HHG signal is estimated in the model by approximating the HHG dipole as $d \propto \rho_0(1-\eta(t))\sqrt{w(\Epsilon)}$ and accounting for on-axis phase matching and absorption as described in \cite{Allison_Thesis2010}. The model can also be easily extended to include the dynamic response of the system to perturbations with the frequency-lock servo loop engaged by adding additional differential equations to the coupled set (\ref{AODE}) \cite{Allison_Supporting}.

%Expressing $f_j = (-\beta_j/2 + i\psi_j)A_j$, the effective single pass loss $\beta(\Delta\omega)$ and phase $\psi(\Delta\omega)$ are the solid and dashed lines, respectively, and use the right scale. The resonance structure observed in $\beta_j$ and $\psi_j$ on the red tail of the spectrum is due to a $\pi$ phase jump in the spectral phase of $A(\Delta\omega)$.

\begin{figure}[t]
\begin{center}
\includegraphics[width = 8.5 cm]{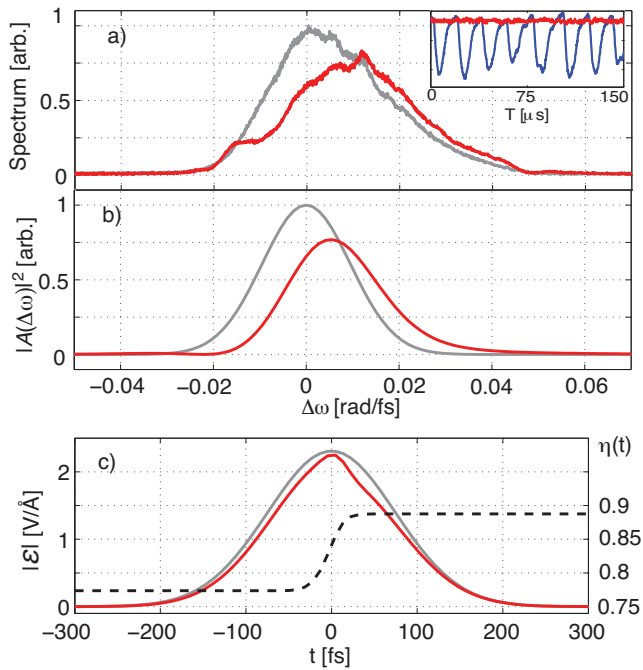} 
\caption{a) Measured intracavity spectra with (red) and without (gray) gas for offset $\Delta \approx 40$ kHz and 38 W pump power. The intracavity power is reduced by 11\%.  b) The high power solution under similar conditions. c) Corresponding time domain $|\Epsilon(t)|$ and $\eta(t)$ (black dashed line, right scale). Inset: Intracavity power vs. time with (red curve) and without (blue curve) the lock offset $\Delta$. With $\Delta = 0$, the system oscillates on the $\sim 20$ $\mu s$ time scales characteristic of the servo loop.} 
\label{tstep_fig}
\end{center}
\end{figure}

With model parameters in the experimentally reasonable ranges $\alpha_1,\alpha_2 \sim 0.2-0.5$, $k_p \sim 0.1-0.2$, $L = 200$ $\mu$m, and $\rho_0 \sim 1-3 \times 10^{17}$ cm$^{-3}$, the model reproduces the measured data. The black dashed curves in Fig. \ref{sweeps_fig} show model results from scanning the $\theta_j$ from negative to positive, which simulates the cavity length scan. The model's error signal and HHG signal are scaled for comparison with the experimental data, but the intracavity power is not. The observed distortion of the cavity resonance line shape can be understood as a new manifestation of self-locking phenomena \cite{Dube_JOSAB1996}, where the non-linear phase shifts here are provided by the intracavity plasma. For large positive $\theta_j$, the systems exhibits optical bistability, with the coupled equations (\ref{AODE}) having two perturbation stable steady-state solutions: a high intracavity power solution in which the plasma phase shifts, with the dominant contribution from $\phi_j$, compensate the cavity length detuning, and a low intracavity power solution with no gas ionization. In the frequency scan, self-locking is lost when the ionization of the medium saturates and the plasma phase shift cannot be increased further, causing the system to jump to the low power solution.
When actively locking the optical frequencies of the comb and cavity, high power operation is found to be perturbation unstable at the resonance condition (error signal = 0), and laser frequency or intensity noise can make the system jump to the low power solution. The frequency-lock servo then returns the system to resonance, producing oscillations as shown in the inset of Fig. \ref{tstep_fig}, with the servo loop effectively scanning through the self-locking curve of Fig. \ref{sweeps_fig} repeatedly. These oscillations reduce the duty cycle and average power of HHG by a factor of $\sim 2$ and are deleterious to the coherence of the XUV comb.

Stable locked operation can be maintained by introducing an offset to the error signal such that the servo aims to maintain a frequency offset $\Delta$ between the shifted resonance condition of the cavity and the pumping comb. However, the intracavity pulse and spectrum are distorted from those of ideal linear operation. Fig. 2a) shows the measured intracavity spectra with and without target gas for $\Delta \approx 40$ kHz and 38 W pump power. The intracavity power is reduced by 11\% and the spectrum is shifted to the blue. More severe problems with servo stability and reductions in the intracavity power occur at higher cavity finesse, reducing the attainable HHG power \cite{Yost_TBP}. Model results under similar conditions are shown in Figs. \ref{tstep_fig}b) and \ref{tstep_fig}c). In the time domain, the beginning of the intracavity pulse maintains better constructive interference with the pump comb than the end of the pulse, which experiences a larger plasma phase shift. In the frequency domain, the self-phase modulation appears as loss on the red side of the spectrum and gain for the blue, and the intracavity spectrum shows a corresponding shift to the blue. Repeating the calculation without the loss term in equation (\ref{Epsapprox}) yields nearly identical results, indicating that self phase modulation is the dominant effect limiting the intracavity power. 
  
\begin{figure}[t]
\begin{center}
\includegraphics[width = 9.0 cm]{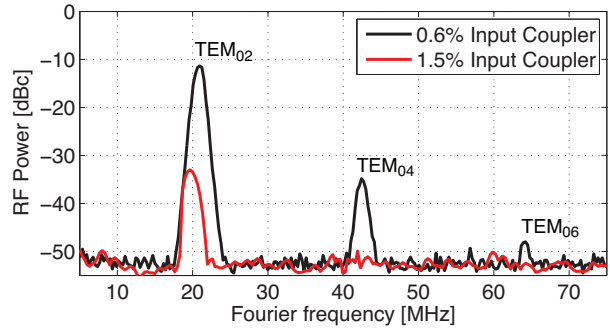} 
\caption{RF spectra of the intracavity power signal with 1 MHz resolution bandwidth. The beat frequencies are labeled with the higher-order modes that have the corresponding frequency offset from the fundamental TEM$_{00}.$}
\label{RF_fig}
\end{center}
\end{figure}

While the 1D model captures the essential physics, plasma lensing effects are present and manifest themselves in the cavity-plasma system in a novel way. The spatial profile of the plasma has higher spatial frequency content than the fundamental TEM$_{00}$ mode of the cavity and can thus couple power into the cavity's higher-order modes. This is clearly observed as modulation of the intracavity power at 21 MHz and its harmonics shown in Fig. \ref{RF_fig}. Inspection of the cavity's higher-order mode structure indicates the modulation comes from beating between the fundamental TEM$_{00}$ mode of the cavity with the progression of higher order modes of even symmetry in the vertical direction. We hypothesize that mode coupling in the vertical direction is preferentially excited due to the smaller spot size and stronger plasma gradient in this dimension. The transition to multimode operation is observed to occur abruptly as the intracavity power and gas flow are increased. While the threshold conditions are typically beyond those found optimal for HHG, it is important to understand and control this effect, as such high frequency modulation severely impacts the phase coherence of the XUV comb. The power coupled into higher order modes and the multi-mode appearance threshold depend dramatically on the cavity finesse, as shown in Fig. \ref{RF_fig}, where increasing the finesse by a factor of $\sim 2.5$ reduced the appearance threshold from 7.6 kW to 5.3 kW and increased the power in higher order modes by more than 2 orders of magnitude.

In conclusion, plasma optical bistability, cavity enhanced spectral blueshifting, and multimode operation induced by plasma lensing must be examined in the design of intracavity HHG systems. Stable locked operation can be maintained in the midst of a large steady-state intracavity plasma by careful servo loop design. Self-phase modulation remains as the primary obstacle to increasing the intracavity power, but modest drops in the finesse can allow large gains in obtainable intracavity power and suppression of multimode operation. Multimode operation can also be suppressed with an intracavity aperture. The optimization of HHG must also be considered, as seen in Fig. \ref{sweeps_fig}, where the HHG signal goes through a maximum while the intracavity power continues to increase. With the current over-coupled cavity design with a finesse of $\sim 400$, we have observed more than 20 $\mu$W out-coupled power levels under stable locked operation in the 13th harmonic. Assuming an outcoupling efficiency of $9\%$ \cite{Yost_OptLett2008}, this corresponds to a generated power of 220 $\mu$W, or a spectral brightness of roughly $4 \times 10^{16}$ photons/(s mm$^2$ mrad$^2$ (0.1\%bandwidth)), approximately one tenth that of the XUV undulators of the Advanced Light Source Synchrotron at this wavelength \cite{ALSweb}. 

This research is funded by DARPA, AFOSR, NIST, and NSF. T. K. Allison and A. Cing\"{o}z are National Research Council postdoctoral fellows. We thank A. Ruehl, M. Fermann, and I. Hartl for developing the high power Yb$^+$ fiber laser.  

Note: Since Submission, we have become aware of related work \cite{Carlson_OptLett2011}

%\bibliographystyle{unsrt}
%\bibliography{/users/tka3/Documents/bibliographies/masterbib}

\end{document}